# Phase diagram and dynamics of the projected SO(5)-symmetric model of high-$T_c$ superconductivity

A. Dorneich[1], W. Hanke[1], E. Arrigoni[1], M. Troyer[2] and S.C. Zhang[3]

[1] *Institut für Theoretische Physik, Universität Würzburg, Am Hubland, 97074 Würzburg, Germany*
[2] *Institut für Theoretische Physik, ETH Zürich, 8093 Zürich, Switzerland*
[3] *Department of Physics, Stanford University, Stanford, 94305 California*

(November 5, 2018)

We present numerical studies of a quantum "projected" SO(5) model which aims at a unifying description of antiferromagnetism and superconductivity in the high-$T_c$ cuprates, while properly taking into account the Mott insulating gap. Our numerical results, obtained by the Quantum Monte Carlo technique of Stochastic Series Expansion, show that this model can give a realistic description of the global phase diagram of the high-$T_c$ superconductors and accounts for many of their physical properties. Moreover, we address the question of dynamic restoring of the SO(5) symmetry at the critical point.

PACS numbers: 74.20.–z, 74.25.Dw, 11.30.Ly, 02.70.Uu

SO(5) theory of superconductivity provides a unifying approach to describe both antiferromagnetism (AF) and superconductivity (SC) in the high-$T_c$ cuprates [1]. Microscopic SO(5)-symmetric models [2,3] successfully describe many features of high-$T_c$ cuprate physics [4,5] and make a number of experimental predictions [6]. However, the cuprates' Mott insulating behavior at half-filling severely challenges the validity of SO(5) theory. SO(5) symmetry requires collective charge-two excitations to have the same (vanishing) mass as collective spin-wave excitations and not the large Coulomb energy gap $U$ of several eV observed in a Mott insulator at half-filling. To overcome this problem a low-energy effective model – the so-called "projected" SO(5) or pSO(5) model – has been introduced [7,8] in which a Gutzwiller constraint of no-double-occupancy is implemented exactly. This is done by projecting out the particle pair excitations (the so-called $\pi^+$ Goldstone mode) and by retaining only the massless magnon and hole-pair modes. The central hypothesis of pSO(5) theory is that this projected model is accurate in describing both the static and dynamic properties of the high-$T_c$ superconductors (HTSC). For the fermionic excitations, a corresponding pSO(5) model has been shown to provide a natural explanation of the interrelation between SC and AF gaps [8].

In this letter, we compare the properties of the pSO(5) model in two dimensions with a variety of salient features of the HTSC such as the global phase diagram and the neutron scattering resonance. The pSO(5) model turns out to give a good semiquantitative description of the SC phase, the behavior of the chemical potential $\mu$ as a function of doping and of the magnetic resonance peak. We identify a jump in the hole pair density $\delta$ in the $\delta(\mu)$ diagram which is very similar to the available experimental data on La$_{2-x}$Sr$_x$CuO$_4$ [9]. The jump signals coexisting hole-rich (metallic) and hole-free regions, the latter of which becomes antiferromagnetic at $T=0$. We study the nature of the coexistence region, identify a tri-critical point where the coexistence line merges into a Kosterlitz-Thouless critical line, and discuss to what extent the SO(5) symmetry can be restored at this point.

All our results were obtained using a numerically exact Quantum Monte Carlo technique, the Stochastic Series Expansion (SSE) [10–12]. SSE has three important features discerning it from other QMC techniques (see [13] for an overview). First, SSE is an exact method without any systematic error (such as the Trotter discretization error). Second, a non-local loop-update mechanism can easily be implemented within SSE [11]. This approach, which is similar to the loop-algorithm [14], allows for much lower autocorrelations times on large systems or at low temperatures than any purely local update could reach [12]. Such a favorable scaling behavior, which makes it possible to simulate systems with several thousand lattice sites, is neccessary for tracing phase transitions of the Kosterlitz-Thouless type, as we encounter here. Third, we have developed a very efficient way to measure arbitrary Green's functions within SSE [12].

We start with a more detailed presentation of the pSO(5) model. In Ref. [3] it has been shown that the low-energy SO(5) excitations on the rung of a ladder can be cast into a picture of 5 hard-core bond bosons: three triplet states ($t^\dagger_{\alpha=2,3,4}$) and particle and hole-pair states ($t^\dagger_p$ and $t^\dagger_h$). As an effective coarse-grained model, this description can be extended to a two-dimensional system, whereby the excitations are now defined on a $2\times 2$ plaquette [7,15]. The projection is implemented by restricting within the Hilbert space with $t_p(x) = 0$. The projected SO(5) Hamiltonian takes the form [7]

$$\hat{H} = \Delta_s \sum_{x,\alpha=2,3,4} t^\dagger_\alpha(x) t_\alpha(x) + (\Delta_c - 2\mu) \sum_x t^\dagger_h(x) t_h(x) \qquad (1)$$
$$- J_s \sum_{<xx'>,\alpha=2,3,4} n_\alpha(x) n_\alpha(x') - J_c \sum_{<xx'>} (t^\dagger_h(x) t_h(x') + h.c.)$$

where $n_\alpha = (t_\alpha + t^\dagger_\alpha)/\sqrt{2}$ are the three components of the Néel order parameter. $\Delta_s$ and $\Delta_c$ (which models the Mott-Hubbard gap $U$) describe the gap energy of the magnon and the hole-pair states at $\mu = 0$, and the $J_s$ and $J_c$ terms stand for the hopping and for creation/destruction processes of these states.

Let us now turn to the numerical results. We choose $J_s = J_c/2$, corresponding to the "SO(5)-symmetric" point determined in a mean-field analysis of the pSO(5) model [7]. $J := J_s$ will be our unit of energy. Since the magnon density rapidly decreases as soon as one enters the superconducting region (see Fig. 1), the precise value of $\Delta_s$ is not crucial for the dynamics of the model or the general structure of the phase diagram in this phase. This holds as long as $\Delta_s \lesssim 4J$, as for larger $\Delta$ there is no ordered phase in mean field. We, thus, choose $\Delta_s = J$, and shift the chemical potential so that $\Delta_c = \Delta_s$. The mean-field result [7] predicts a phase transition from AF to SC accompanied by a jump in the hole-pair and magnon densities at $\mu_c = 0$, the state at $\mu = \mu_c$ being a coherently mixed AF+SC phase without phase separation. However, Gaussian fluctuations change the picture and predict a first-order transition. Here, we want to study this region with an appropriate, i.e. exact numerical evaluation. In fact, we expect the picture to change appreciably, since no long-range order is allowed in two dimensions.

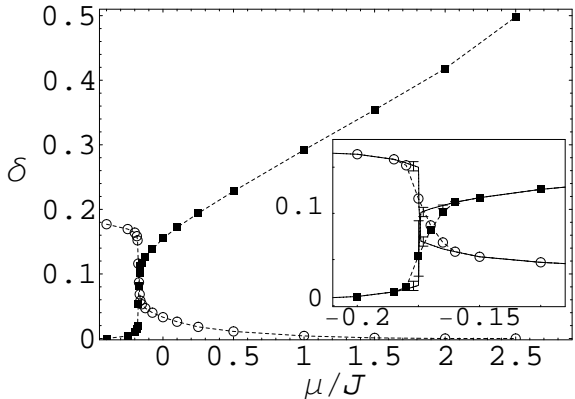

FIG. 1. Hole concentration $\delta = \rho/2 = \frac{1}{2}\langle t_h^\dagger t_h \rangle$ (filled squares) and magnon density $\sum_\alpha \frac{1}{2}\langle t_\alpha^\dagger t_\alpha \rangle$ (circles) as a function of the chemical potential $\mu$ at $T/J = 0.03$. The plotted points result from a finite-size scaling with lattice sizes $V = 10 \times 10$, $14 \times 14$ and $20 \times 20$. The small inlay shows a detailed view to the $\mu$ region in which the hole-pair density jumps to a finite value. The additional solid lines with error bars are $T = 0$ data obtained from a simulateneous scaling of $\beta \to \infty$ and $V \to \infty$ (with lattice sizes of $V = 8 \times 8$, $10 \times 10$, $12 \times 12$, $16 \times 16$, $20 \times 20$ and $\beta = 4.8, 7.5, 10.8, 19.6$, and $30$).

In Fig. 1 we plot the mean hole-pair and magnon densities as a function of the chemical potential for $T/J = 0.03$ and their $T \to 0$ extrapolations. The jump can be clearly seen at $\mu_c = -0.175$, shifted with respect to the mean-field value due to the stronger fluctuations of hole pairs as seen in the Gaussian contributions [7]. The size of the jump, $\delta_c \approx 0.11$, as well as as its slope for $\mu > \mu_c$, $\partial\delta/\partial\mu \approx 0.13$, are very close to the mean-field values of $1/8$ [16]. Notice that, while the hole-pair density rapidly vanishes below $\mu_c$, a nonvanishing magnon density persists well beyond $\mu_c$ even at $T = 0$, due to the fact that the magnon number is not conserved.

A jump in the density as a function of $\mu$ has also been measured in La$_{2-x}$Sr$_x$CuO$_4$ [9]. A comparison with the experimental data of Ref. [9] is displayed in Fig. 2 [16]. As one can see, our data reproduce the experimental results within error accuracy. The data are fitted by adjusting the energy scale $J$ to 220 meV. This value of $J$ has the correct order of magnitude for a typical magnetic superexchange interaction in the cuprates [15].

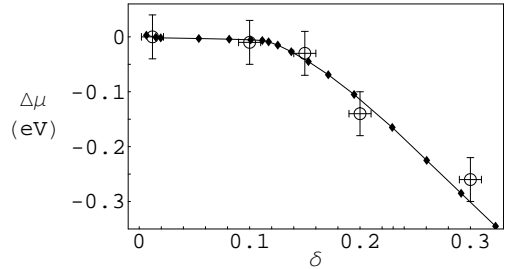

FIG. 2. Chemical potential shift $\Delta\mu$ as a function of hole concentration $\delta$ (open circles) for La$_{2-x}$Sr$_x$CuO$_4$ at $T = 77$ K. The diamonds and the solid line show numerical data for the infinite lattice obtained by finite-size scaling with $J = 220$ meV and $T/J = 0.03$, which corresponds to $T \approx 77$ K.

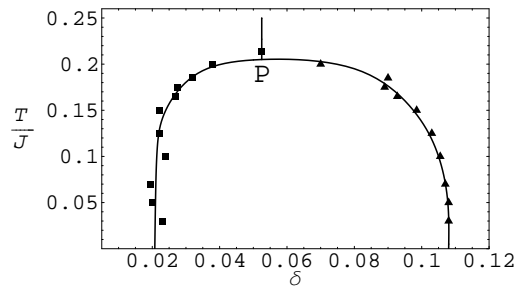

FIG. 3. Hole densities of the coexisting phases on the first order transition line from (almost) zero to finite hole density at $\mu = \mu_c$ as a function of temperature.

The nature of the phase transition at $\mu = \mu_c$ can be determined by studying histograms of the hole-pair distribution for fixed $\mu = \mu_c$. While in an homogeneous phase the density is peaked about its mean value, at $\mu = \mu_c$ we obtain two peaks indicating a first-order transition with a phase separation between (almost) hole-free regions and regions with high hole-pair density. From Fig. 3 we see that the transition is first order for $T < T_P = (0.20 \pm 0.01)J$ at $\mu = \mu_P = (-0.168 \pm 0.002)J$. Above $T_P$, the histograms show strongly fluctuating hole-pair densities suggesting the presence of critical behavior. Indeed, as we will show below, the Kosterlitz-Thouless transition line between the SC and the paramagnetic phase merges at the point P, which is thus a tricritical point.

In realistic electron systems, however, long-range Coulomb repulsion between the doubly-charged hole pairs should heavily disfavor phase separation. To study the effect of off-site Coulomb interaction we have added additional nearest-neighbor and next-nearest-neighbor Coulomb repulsions $V$ and $V' = 0.67\,V$ to the pSO(5) model. Indeed a relatively moderate Coulomb repulsion of $V/J \approx 0.2$ is enough to completely destroy the phase separation. The interesting effect of Coulomb interaction in two dimensions is, thus, to push down the tricritical point into a quantum-critical point at $T=0$. This quantum-critical point now separates the SC from the AF phase and could be a possible candidate for the dynamic restoring of SO(5) symmetry.

The next question concerns the nature of the phase above the tricritical point (or above the quantum critical point in the presence of Coulomb repulsion): do we find microscopic density modulations in the form of static or dynamic stripes or a phase which is homogeneous also on the microscopic level? Careful studies of the static density-density correlation $\langle n_h(r)\, n_h(0)\rangle$ and the dynamical response function $D(\mathbf{k},\omega) = \langle n_h(\mathbf{k},\omega)\, n_h(\mathbf{k},0)\rangle$ show that there is no sign of stripe formation, neither without nor with the above (moderate) Coulomb repulsion.

We have anticipated the existence of a SC phase for $\mu > \mu_c$. In fact, at $T>0$ a true long-range order is prohibited by the Mermin-Wagner theorem. However, we can still have a KT phase of finite superfluid density $\rho_s$ at finite temperature, which is identified by a power-law decay of the SC correlation function

$$C_{h(r)} = \left(t_{h(r)}^\dagger + t_{h(r)}\right)\left(t_{h(0)}^\dagger + t_{h(0)}\right).$$

The transition separates long-range power-law ($C_h(r) \propto r^{-\alpha}$) from rapid exponential decay ($C_h(r) \propto e^{-\lambda r}$). A reliable and accurate distinction between these two decay behaviors requires a finite-size scaling with large system sizes, as well as an efficient QMC estimator for the Green functions appearing in the correlation function. With its non-local update scheme and with our new estimators for arbitrary Green functions SSE provides both. An alternative method for detecting a KT transition exploits the fact that the superfluid density jumps from zero to a finite value at the KT temperature $T_{KT}$ [17]. Since within SSE the superfluid density can be measured quite easily by counting winding numbers [18], this criterion is numerically even more favorable than the arduous direct determination of decay coefficients.

Fig. 4 plots the phase diagram obtained by applying both criteria independently. The figure shows that the projected SO(5) model indeed has a KT phase with quasi long range order whose form in $\mu$-$T$ space looks like the one of the high-$T_c$ cuprates. Both criteria result in exactly the same clearly pronounced phase separation line. It is well known that a similar transition cannot occur for antiferromagnets [19], and that the finite-$T$ AF correlation length $\xi$ is always finite and behaves like $\xi \propto e^{2\pi\rho_s/k_B T}$, $\rho_s$ being the spin stiffness. This fact is confirmed by our numerical results.

In the preceeding paragraphs we have presented some general properties of the pSO(5) model. The question "how SO(5)-symmetric is the pSO(5) model", i.e. the question whether there exists a point on which the full SO(5)-symmetry is dynamically restored, still remains open. As one can see from Eq. (1), the excitation energy for hole pairs can be compensated by $\mu$, in order to have equal energies for *local* spin and hole-pair excitations. Due to this partial compensation the *mean-field* ground state of this model recovers exact SO(5) invariance at $J_c = 2\,J_s$ and $\Delta_s = \Delta_c$ [7]. However, since the Casimir operator of the SO(5) group does not commute with the Hamiltonian, this invariance is not exact, and a symmetry breaking effect can already be seen at the Gaussian level [7]. For a *classical* three-dimensional SO(5)-symmetric model, numerical simulations indicate that the symmetry is asymptotically restored at a bicritical point provided the symmetry-breaking terms have the appropriate sign [5,20]. This is in contrast with the prediction from the $\epsilon$-expansion [21], which would suggest a fluctuation-induced first-order transition. On the other hand, other arguments seem to indicate that the decoupled fixed point is slightly attractive [22]. This discrepancy clearly indicates that *strong-coupling* effects play an important role, calling for an appropriate, i.e. numerically exact treatment, as it is provided by SSE.

One neccessary condition for an SO(5)-symmetric point is that the formation energies of hole-pair bosons and of magnons are identical. This condition is fulfilled along the line from $S$ to the tricritical point $P$ in Fig. 4. Another neccessary condition is that hole pairs and magnons behave in the same way at long distances. This condition is fulfilled on the dashed line in Fig. 4, where the AF and SC correlation lengths $\xi$ become equal. Interestingly, these two conditions meet (within error bar accuracy) at the tricritical point P. Of course, here the correlation length is still finite; however, we find relatively large $\xi$ values of order 10 to 15 in the immediate vicinity of point P, demonstrating the importance of SO(5) critical fluctuations in this region.

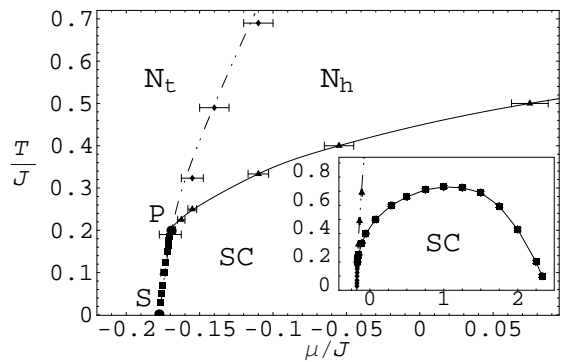

FIG. 4. Phase diagram of the pSO(5) model: The squares between S and the tricritical point P trace the first-order line of phase separation. The solid line from P to the right edge of the plot traces the Kosterlitz-Thouless transition between the "SC" and the normal state. The dashed line separating $N_t$ (=triplet dominated region) and $N_h$ (=hole pair dominated region) describes the line of equal AF and SC correlation lengths. The small inlay shows the same phase diagram on a larger $\mu$ scale, covering the whole "SC" KT phase.

One of the main features of SO(5) theory is that it provides an elegant explanation for the neutron resonance peak observed in some high-$T_c$ cuprates at $k=(\pi,\pi)$ [1]. Experiments show that the resonance energy $\omega_{\rm res}$ is an increasing function of $T_c$, i.e. $\omega_{\rm res}$ increases as a function of doping in the underdoped and decreases in the overdoped region [23]. Fig. 5 shows the corresponding spin correlation spectrum obtained from the projected SO(5) model. The "AF" and underdoped SC regions are described correctly: spin-wave excitations are massless Goldstone modes in the AF phase at $\mu < \mu_c$ (and $T=0$), and become massive when entering into the SC phase. $\omega_{\rm res}$ increases monotonically up to optimal doping $\mu_{\rm opt} \approx 1$. In the overdoped range, however, $\omega_{\rm res}$ is further increasing in contrast to what happens in the cuprates. This is not surprising given the fact that SO(5) theory was developed to model the interplay of antiferromagnetism and superconductivity in the vicinity of the AF-SC transition, i.e. in the underdoped range. The resonance peak continuously looses weight when increasing $\mu$, which is consistent with experimental observations [23].

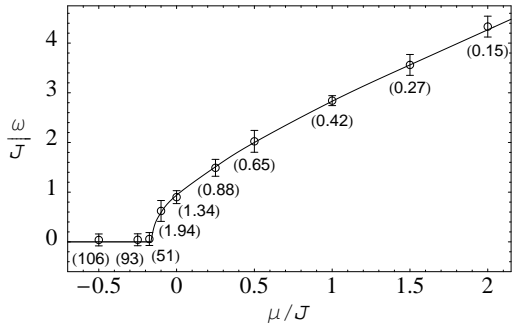

FIG. 5. Dispersion of the $(\pi,\pi)$ peak of spin correlation as a function of the chemical potential The numbers in parentheses indicate the peak weights, i.e. the area under the peak. ($20 \times 20$ lattice at temperature $T/J = 0.1$).

A comparison of the critical temperature $T_c$ obtained from Fig. 4 and $\omega_{\rm res}$ at optimal doping yields the ratio $T_c/\omega_{\rm res,opt} = 0.23$. This is again in accordance with the corresonding ratio for YBa$_2$Cu$_3$O$_{6+x}$, for which the experimentally determined values $T_c = 93$ K (thus $k_B T_C = 8.02$ meV) and $\omega_{\rm res,opt} = 41$ meV yield $T_c/\omega_{\rm res,opt} = 0.20$.

In summary, we have shown that the projected SO(5) model – which is the simplest bosonic model containing two key ingredients for the HTSC: the Mott gap and the vicinity of AF and SC phases – gives a semiquantitative description of many properties of the HTSC in a consistent way. In particular, we have identified a KT-SC phase whose bounds look similar to the the ones of real cuprate materials. Also the doping dependence of the chemical potential as well as that of the neutron resonance peak in the underdoped regime are reproduced correctly. Although the second issue discussed in this letter, the dynamic restoring of SO(5) symmetry in the vicinity of a critical point, cannot be conclusively answered in $D=2$ dimensions, we have found a critical point scenario which makes a *partial* dynamic restoring of SO(5) symmetry at a $T \neq 0$ bicritical point in $D=3$ a good possibility. This will be discussed in a future publication.

The authors acknowledge financial support from BMBF (05SB8WWA1), DFG (HA 1537/16-1,2 and HA 1537/17-1), KONWIHR OOPCV and the Swiss National Science Foundation. The calculations were carried out on a Cray T3E at the high-performance computing center HLRZ (Jülich).

[1] S.C. Zhang, Science **275**, 1089 (1997).
[2] D. Scalapino, S.C. Zhang and W. Hanke, Phys. Rev. B **58**, 443 (1998).
[3] R. Eder *et al.*, Phys. Rev. B **59**, 561 (1999); E. Arrigoni and W. Hanke, Phys. Rev. Lett. **82**, 2115 (1999).
[4] E. Demler, H. Kohno and S.C. Zhang, Phys. Rev. B **58**, 5719 (1998);
[5] X. Hu, cond-mat/0011203.v2 (unpublished);
[6] D. Arovas, *et al.*, Phys. Rev. Lett. **79**, 2871 (1997); E. Demler *et al.*, Phys. Rev. Lett. **80**, 2917 (1998); P.M. Goldbart and D.E. Sheehy, Phys. Rev. B **58**, 5731 (1998).
[7] S.C. Zhang, J.P. Hu *et al.*, Phys. Rev. B **60**, 13070 (1999).
[8] M. G. Zacher, W. Hanke *et al.*, Phys. Rev. Lett. **85**, 824 (2000); E. Arrigoni, M. G. Zacher and W. Hanke, cond-mat/0105125 (unpublished).
[9] A. Ino *et al.*, Phys. Rev. Lett. **79**, 2101 (1997).
[10] A. W. Sandvik, Phys. Rev. B **56**, 11678 (1997).
[11] A. W. Sandvik, Phys. Rev. B **59**, R14157 (1999).
[12] A. Dorneich, M. Troyer, cond-mat/0106471 (unpublished).
[13] W. von der Linden, Phys. Rep. **220**, 53 (1992).
[14] H. G. Evertz *et al.*, Phys. Rev. Lett. **70**, 875 (1993).
[15] Here, however, we are assuming a "phenomenological" pSO(5) model, rather than evaluating the coarse-grained parameters from some microscopic Hamiltonian, like, e.g., the Hubbard or t-J model (see Ref. [8] and E. Altman, A. Auerbach, unpublished).
[16] Notice that the "physical" doping $\delta$ is equal to $\rho/2$, due to the fact that $t_h^\dagger$ creates a hole pair in a $2 \times 2$ plaquette.
[17] D. R. Nelson and J. M. Kosterlitz, Phys. Rev. Lett. **39**, 1201 (1977).
[18] K. Harada and N. Kawashima, Phys. Rev. B **55**, R11949 (1997).

[19] S. Chakravarty, B.I. Halperin and D.R. Nelson, Phys. Rev. Lett. **60**, 1057 (1988) and Phys. Rev. B **39**, 2344 (1989).
[20] Interestingly, these terms are precisely the ones originating from quantum fluctuations due to the projection, as obtained from a *weak-coupling* expansion in E. Arrigoni and W. Hanke, Phys. Rev. B **62**, 11770 (2000).
[21] J. M. Kosterlitz, D. R. Nelson, and M. E. Fisher, Phys. Rev. B **13**, 412 (1976).
[22] A. Aharony, cond-mat/0107585
[23] H.F. Fong *et al.*, Phys. Rev. B **61**, 14773 (2000); H. He *et al.*, Phys. Rev. Lett. **86**, 1610 (2001).